# How well can machine learning predict demographics of social media users?

Nina Cesare (ninac2@uw.edu),[1] Christan Grant (cgrant@ou.edu),[2] Quynh Nguyen (quynh.ctn@gmail.com),[3] Hedwig Lee (hedwig.lee@wustl.edu),[4] Elaine O. Nsoesie (en22@uw.edu)[1*]

1. Institute for Health Metrics and Evaluation, University of Washington, Seattle, WA 98121, USA
2. School of Computer Science, University of Oklahoma, Norman, OK 73019, USA
3. Department of Epidemiology and Biostatistics, University of Maryland College Park, MD 20742, USA
4. Department of Sociology, Washington University, St. Louis, MO 63130, USA

*Corresponding author

## ABSTRACT

The wide use of social media sites and other digital technologies have resulted in an unprecedented availability of digital data that are being used to study human behavior across research domains. Although unsolicited opinions and sentiments are available on these platforms, demographic details are usually missing. Demographic information is pertinent in fields such as demography and public health, where significant differences can exist across sex, racial and socioeconomic groups. In an attempt to address this shortcoming, a number of academic studies have proposed methods for inferring the demographics of social media users using details such as names, usernames, and network characteristics. Gender is the easiest trait to accurately infer, with measures of accuracy higher than 90% in some studies. Race, ethnicity and age tend to be more challenging to predict for a variety of reasons including the novelty of social media to certain age groups and a lack of significant deviations in user details across racial/ethnic groups. Although the endeavor to predict user' demographics is plagued with ethical questions regarding privacy and data ownership, knowing the demographics in a data sample can aid in addressing issues of bias and population representation, so that existing societal inequalities are not exacerbated.

## INTRODUCTION

Social media platforms offer (near) real-time unsolicited information on users' thoughts, feelings and experiences, allowing researchers to track attitudes and behaviors as they emerge and over time. The individual and population scale of these data enables the study of the flow of information through complex networks that are difficult to assess using traditional means of data collection [1]. These data are also usually low-cost compared to traditional modes of data collection, such as surveys.

Researchers have used these data to study social processes across a wide domain, ranging from economics to public health. For example, posts on Facebook and Twitter have been used in real-time tracking of public opinion on political events and social issues [2–7]. In public health, these data have been used to study attitudes toward tobacco use [8–11], vaccinations [12, 13], sleep disorders [14], mental health [15, 16], patient-perceived quality of care in hospitals [17], food deserts [18], foodborne illness [19–22], trends in physical activity, weight loss, and diet [23–25], and the dynamics of infectious diseases (such as, influenza and dengue) [26–32].

Studies have also highlighted potential limitations and biases of using these data for public health [29, 33, 34]. One such limitation is a lack of demographic detail in many digital data sources which limits their applicability for research and practice in fields such as demography and public health, where it is pertinent. For example, demographic information can enable the study of health disparities between populations from different racial or ethnic backgrounds, sex, age and socioeconomic status to ensure that the use of these data in public health practice does not increase existing health inequalities. Knowledge of demographics can also allow for procedures to reweight the data to help adjust for over- and under-representation from certain groups. This lack of demographic details has led to the proposal of a range of methods for inferring the demographics of social media users based on their profile, postings, and network information. Studies typically assume that a user's color preference, name, username, network characteristics, language usage and other attributes can reveal demographic traits such as, age, gender and race or ethnicity. While targeted advertising and recommender systems on the Internet suggest that advanced demographic prediction methods exist, the massive troves of data used in these processes are usually not available to academicians.

Here, we perform a comprehensive review on the performance of approaches proposed to detect the gender, age, race and ethnicity of social media and blog users in the academic literature. We characterize trends in the literature, discuss how the methods and metadata used in these studies could impact predictions, and identify open areas for development.

## Selection and Evaluation of Studies

We conducted an initial literature search in November 2016 and updated the search in November 2017. We queried combinations of the following term groups: "demographic, gender, age, or race", "prediction, classification, or machine learning" and "social media, Twitter or blogs". We searched three databases – Engineering Village, Microsoft Academic, and the ACM Digital Archives – since preliminary results indicated most of the papers were published in computer science conference proceedings. We supplemented these searches with findings from Google Scholar, which provides access to a broad scope of academic literature, including conference proceedings and pre-prints.

In total, we identified 160 potentially relevant articles based on their titles and summaries or abstracts. We focused on articles that aimed to infer or predict the demographics of social media and blog users from information provided within that context. We therefore excluded studies that predict demographic attributes based on other sources of digital data, such as browsing history or mobile phone messages. Additional exclusion criteria were applied to arrive at 91 articles fitting the focus of this review (see Figure 1). We discuss the methods, group and present studies based on the demographic trait predicted.

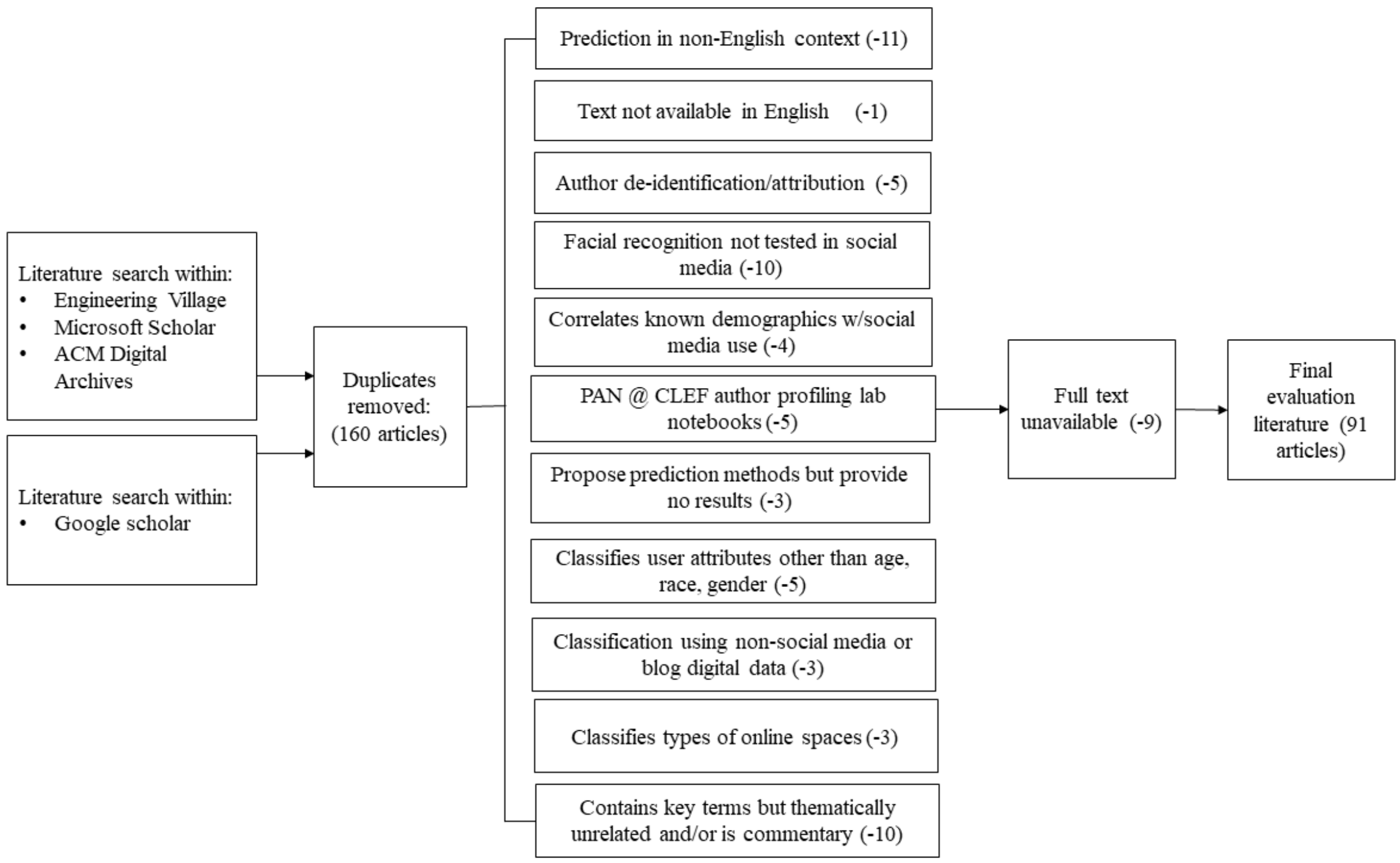

**Figure 1: Literature identification and selection.** We selected 160 potentially relevant articles based on title and abstract from three academic databases and Google scholar. After excluding irrelevant studies, we arrived at 91 articles that proposed methods for inferring the age, race, ethnicity or gender of social media users.

## RESULTS

### Summary of Methods Used in Prediction

The methods presented in the selected manuscripts included machine learning approaches, raw and adjusted data matching approaches, and human and automated facial recognition tools. Seventy-five (82.4%) studies used supervised machine learning approaches, while two studies used unsupervised learning. The most common machine learning methods considered included, support vector machines [35–56], naïve Bayes [57–62], variants of the winnow algorithm [63–66], k-nearest neighbors [37, 67, 68], naïve Bayes decision tree hybrid [69, 70], gradient boosted decision trees [71, 72], neural networks [73–75], random forests [76], entropy models [51, 61, 77], and standard and regularized linear, logistic or log-linear regression [78–92]. These methods were applied to data from various platforms (see Figure 2), with an overwhelming focus on Twitter (n=57, 62.6%), likely due to the availability and wide use of Twitter data for research.

Five (5.5%) studies inferred demographics based on facial evaluations from either paid human evaluators or facial recognition algorithms. The remaining 11 (12.1%) studies used raw or adjusted data matching, meaning they directly linked user data to information from sources such as the United States Census Bureau, or estimated the probability of belonging to particular demographic categories given a combination of known demographic information (e.g., prior data on the popularity of first names over time) and information from users' profiles.

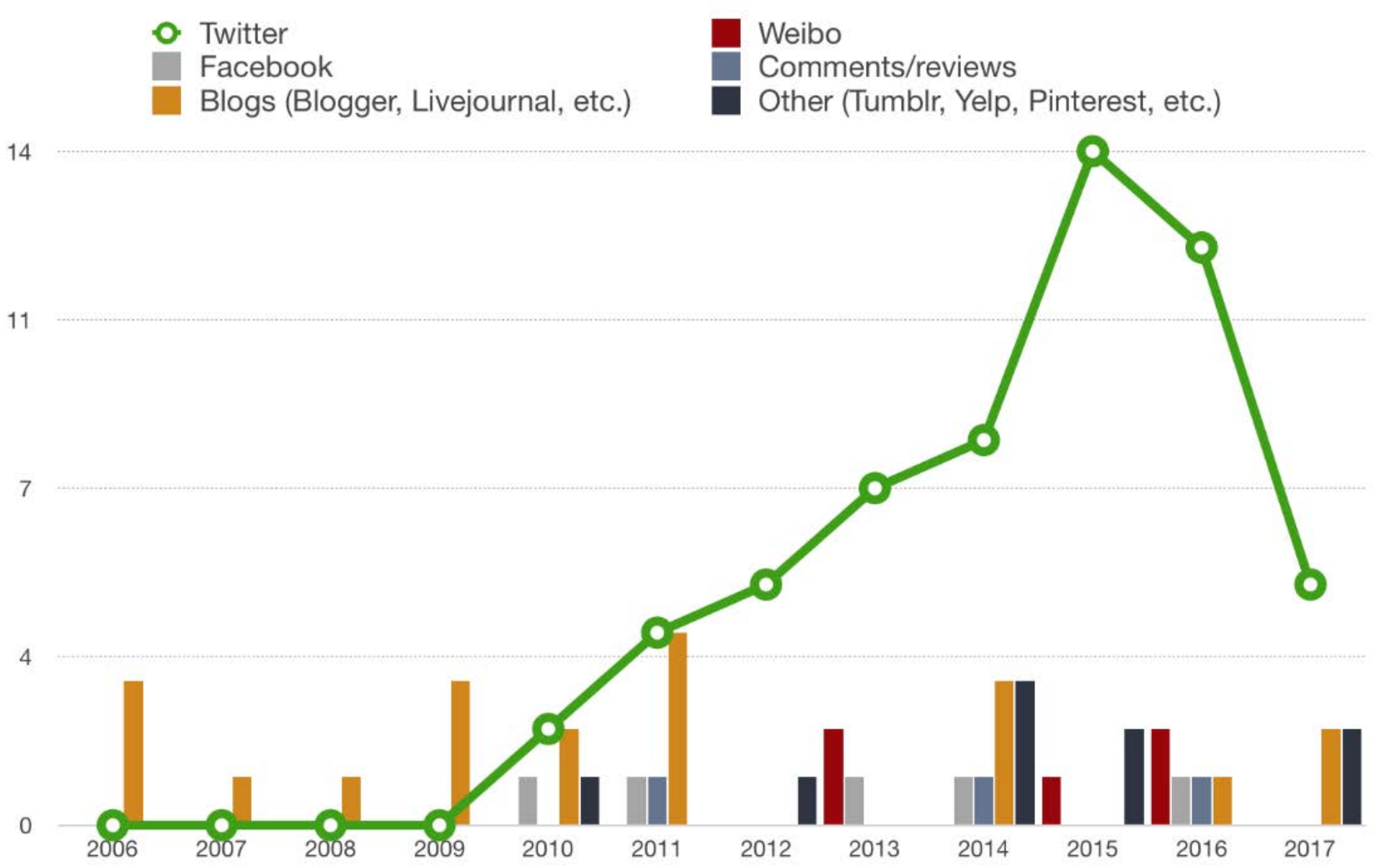

**Figure 2: Literature by year.** Early predictions of user demographics focused on Blogs. However, research predicting the traits of Twitter users is most common.

In addition, the measures used to assess the performance of the proposed algorithms included, *accuracy* [(true positives + true negative)/total population], *precision* (also known as positive predictive value = true positives/predicted positive condition), *recall* (also known as sensitivity= [true positive/(true positives + false negatives)], *F1-score* (the harmonic mean of the precision and recall), and *area under the curve (AUC)* (plot representing sensitivity/specificity pairs corresponding to different decision thresholds; is a measure of how well a parameter can distinguish between two diagnostic categories). Specifically, 62 (87.3%), 37 (84.1%) and 12 (60.0%) of the gender, age and race/ethnicity prediction approaches respectively, reported findings using a specific performance metric. Accuracy was the most commonly reported metric; reported in 74.6%, 47.7%, and 35.0% of studies on gender, age and race/ethnicity prediction studies respectively. The distribution of accuracy by trait is shown in Figure 3. Studies predicting gender had the highest median accuracy, with some studies reaching scores higher than 0.90. Although all studies on age prediction that report accuracy used the same methodological approach, the accuracy fell across a broad range (0.43, 0.95). In contrast, the prediction accuracy for race/ethnicity fell within a narrower range and had a median value of approximately 0.81.

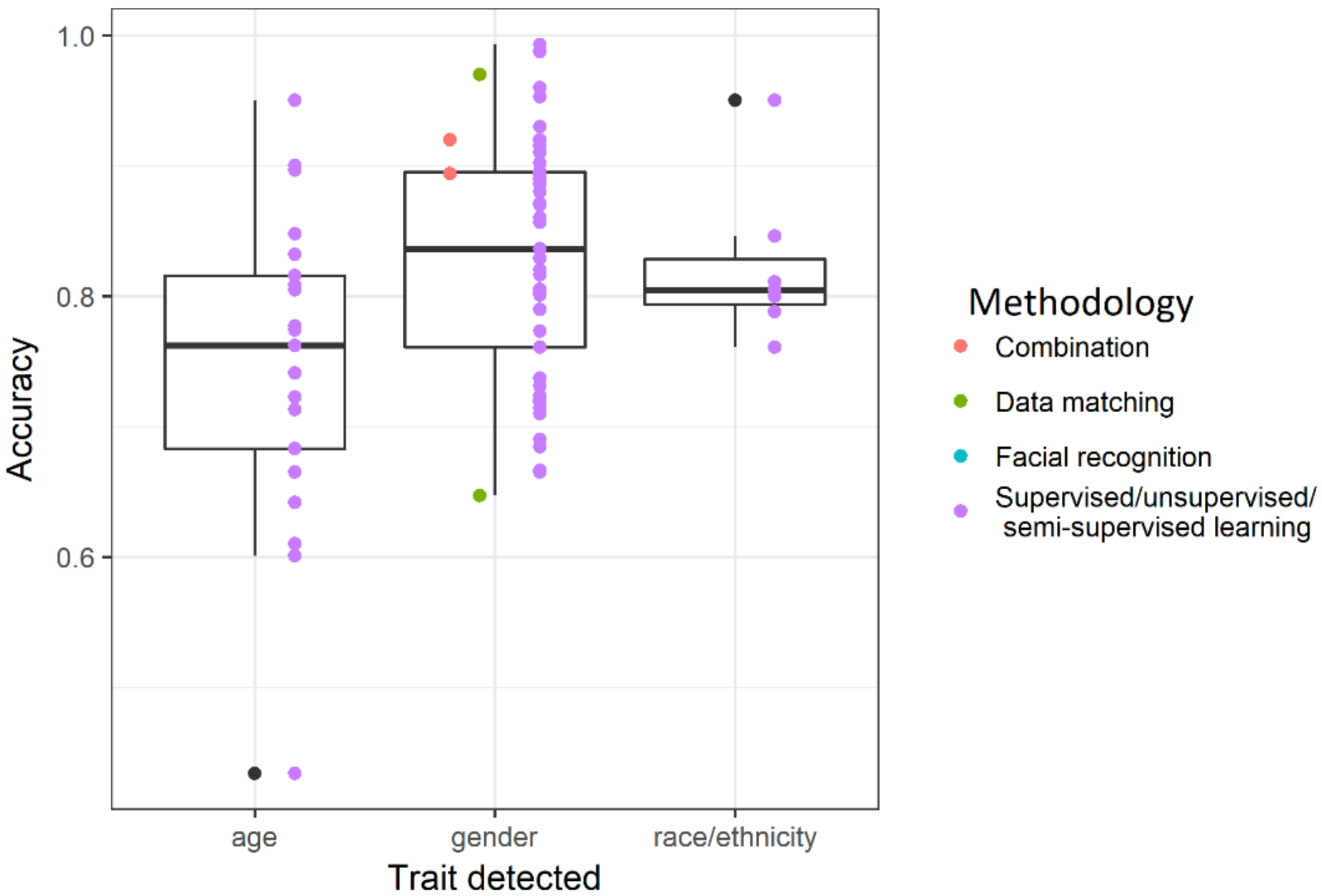

**Figure 3: Distribution of Predicted Accuracy by Trait.** Methods predicting gender perform best. The median accuracy of methods used in inferring age is similar to that of gender, but the distribution varies widely. In contrast, the range of accuracies observed in the prediction of race and ethnicity is narrower.

**Predicting Gender**

Seventy-one (78.0%) studies focused on gender prediction [35, 37–40, 42–45, 47–49, 51–53, 57–70, 73, 74, 77–86, 91, 93–121]. Of these, 54.9% employed user postings, 25.3% used information from user profiles (such as, profile colors), and the remainder used a combination of postings, network and profile information.

Most studies proposing classification methods focused on linguistic patterns in names and posted text. Studies that employed users' names either linked first names to offline survey data or used these data to build models designed to improve predictive algorithms that incorporate other characteristics. Although some sites (e.g., Twitter) do not require users to provide a valid first and last name in their profile, this technique may still capture a large sample of users. For example, Mislove et al. [108] found that 64.2% of Twitter users elect to include a first name in their profile. However, approximately 71.8% of these Twitter users were male.

The performance metrics were consistently above 0.50. Of the studies that used accuracy (n=53, 74.6%) as the measure of evaluation, the average accuracy was 0.83 (range 0.65 to 0.99). Studies that reported precision also had high values, with an average of 0.84 (range 0.58 to 0.98). Reported F1-scores were also high, with an average value of 0.84 (range 0.68 to 0.98). We excluded the values reported by Miller et al. [59] because they reported results for females but not males. Overall, most of the proposed algorithms could reliably predict the gender of social media users.

Bamman et al. [101] suggest that gender is heavily embedded within individuals' identities. It may be equally embedded within social media users' online personas to the extent that unsupervised clustering techniques can differentiate between users with distinct gender identities. For example, Alowibdi et al. [69] found that profile colors alone can be effectively used to infer the gender of Twitter users'. They predicted the gender of Twitter users with 71.4% accuracy using only five quantized colors (i.e., compression via the reduction in the number of colors) from users' profile fields.

Another reason why gender lends itself well to profile-based prediction techniques is because a person's name can be a strong predictor of gender. For instance, Liu and Ruths [42] found that adding names data to a text-dependent gender prediction improved the accuracy of their method from 0.83 to 0.87. Both Fernandez et al. [74] and Karimi et al. [113] suggested inferring gender by combining user names with results from facial recognition software applied to profile photos. Mueller and Stumme [44] and Vicente et al. [105] found that even if a name is not directly identifiable, it can either be *cleaned* in a way that renders the true name linkable with official census data (for example, converting "Hi I'm Kaylaaaaa" to "Kayla"), or it may contain extractable features that are gender dependent (e.g. a name that ends in a vowel is more likely to imply a female user than a male user). Furthermore, Cesare et al. [122] also developed a weighted ensemble model that used word and character n-grams from user names [65], the linguistic structure of names [44], and direct matches between user names and U.S. Social Security Administration (SSA) data to obtain an accuracy of 0.83, precision of 0.82, recall of 0.85, and F1-score of 0.84.

Studies that solely used profile metadata or combined it with users' posts had higher accuracies than those that solely use information from users' postings. This suggests that simple, subtle cues embedded within users' profiles can be used to accurately detect gender. In addition, studies that used non-algorithmic methods (i.e. simple or adjusted data matching) also performed comparably well as studies that used machine learning algorithms (i.e., average accuracy of 0.81 and 0.82 for data matching and machine learning, respectively)

**Predicting Age**

We identified 44 (48.4%) studies that predicted age [35, 37, 49, 52, 53, 56, 58, 62–64, 73, 75, 76, 78–80, 83, 85, 87, 89–98, 104, 106, 110, 111, 117, 118, 120, 123–130]. The proposed algorithms were overwhelmingly text-dependent. These studies used text features and supervised learning methods to predict exact numeric age, age category (adolescent/adult/senior) and life stage (e.g., student, employed, retired). Most studies capped predicted age between 30 and 40, perhaps due to the fact that social media users tend to be younger. The reported average F1-scores of studies that capped predicted age in the 40s was comparable to that of studies that capped predicted age in the 30s (F1=0.74 and F1=0.71, respectively).

Several studies suggested that profile photos may also be used to assess age. McCormick et al. [110] found that human coders are highly reliable when assessing a user's age range. When shown 1,000 Twitter user profile images, at least two out of three workers on Amazon's Mechanical Turk could identify the numeric age range of 96.6% and the life stage of 97.9% of users. An and Weber [93] used Face++, a free facial recognition service that can estimate a user's age within a 10-year span and observed that 49% of the 346,050 Twitter users in their dataset had a detectable face, which is low. When comparing the predicted age of the user and the age extracted from their profile description, they found that 25% of the predictions were within a three-year difference and 75% of predictions were within an 11-year difference of the user's extracted age. However, these studies did not compare estimates to ground truth data (i.e., true demographic information about the users sampled). Rather, estimates were compared to population values or data not drawn from the population studied. Further work is needed to assess the reliability, and best applications of facial recognition by human and automated coders.

Simple approaches – such as identifying mentions of age within a user's profile or matching first names to trends in popularity over times – could be useful but not widely employed within existing literature beyond the development of a training dataset. For the former, the available data might be limited to teenagers or milestone years, and the latter may provide reliable predictions for popular names only.

The reported high accuracy observed in age prediction suggests that clues from users' postings - such as, the sociolinguistic indicators or topics discussed - provide valuable information about the age of the user. The accuracy and F1-scores of age prediction studies that relied exclusively on information from users' postings were slightly higher than those of studies that combined data from users' postings with profile or network information (accuracy=0.76, 0.72, respectively; F1-score=0.80, 0.74, respectively). Thus, using additional data besides profiles posts did not significantly improve prediction of age.

These studies indicate that age prediction is more difficult with additional categories. In general, the performance of prediction algorithms decreased as the number of predicted categories increased. The average F1-scores of studies that predicted binary age categories (e.g. under/over 20 years old), three age categories (e.g. under 20, 20-40, above 40), and four age categories (e.g. 10-19, 20-29, 30-39, 40+) were 0.91, 0.73 and 0.71, respectively. The few studies that predicted numeric age, reported an average correlation of 0.73. While prediction of numeric age is challenging, it is most beneficial since the results can be binned into different categories as needed.

**Predicting Race and Ethnicity**

Twenty (21.9%) of the selected studies addressed the prediction of race or ethnicity [48, 49, 54, 71, 72, 78, 79, 82, 85, 86, 91, 93, 107–111, 126, 131, 132]. The earliest studies were from 2010, compared to 2006 for age and gender, likely due to the challenge of predicting this demographic trait. Thirteen of these studies reported an average accuracy of 0.82 and twelve studies reported an average F1-score of 0.70. Eight studies did not report a performance metric.

Nine (45.0%) race/ethnicity prediction studies solely relied on users' postings or combined users' postings with profile and network metadata to infer race. The performance of these studies suggests that posted text provides valuable insight into the race or ethnicity of the user, but that performance improves considerably when these data are analyzed with the added context of profile or network metadata. The average accuracy and F1-scores of text-only methods was 0.76 and 0.51, respectively, while the average accuracy and F1-scores of methods that combined text with additional metadata was 0.81 and 0.73, respectively.

Six studies solely used profile metadata to predict race. In McCormick et al. [110], human analysts labeled race based on profile photos. At least two out of three human coders agreed on the race of 95.5% of 1,000 user photos. Chen et al. [49] applied n-grams and Linear Discriminant Analysis (LDA) to data on Twitter users' neighbors' descriptions to predict race with 78.8% accuracy (F1=0.75).

Additionally, last name was a common feature used in studies that employed profile data alone or in conjunction with other metadata. Longley et al. [111] used the Onomap system – an online index of names linked to geographic information developed by Mateos and colleagues [133] – to infer the ethnicity of Twitter users in London. Similarly, Hofstra et al. [107] used last names to determine whether Dutch Facebook users were Dutch, Moroccan or Caribbean. Neither Longley et al. [111] or Hofstra et al. [107], however, reported performance metrics given ground truth data. Rao et al. [109] illustrated that last names provide valuable information that can be used solely or in conjunction with user postings to predict the ethnicity of Nigerian Facebook users with an accuracy of approximately 0.81. Similar to the use of first names for age prediction, ethnicity is commonly tied to last name making it traceable and useful for prediction.

Similarly, Chang et al. [131] and Mislove et al. [108] applied a method for inferring the race of U.S. Twitter users based on last names and the aggregate racial distribution of the population from which the user is drawn. However, they failed to assess their prediction against individual ground truth data. Additional work is needed to rigorously assess the potential of this prediction technique.

## DISCUSSION

We performed a comprehensive review of studies that propose methods for inferring the age, race, ethnicity and gender of social media and blog users. We examined trends in the literature by trait and assessed the performance of the proposed methods.

We note that gender was the easiest trait to infer using a variety of methods and features. In contrast, in the prediction of age, methods that relied on data from users' postings reported higher accuracies than methods that did not use data from users' postings or combined these data with other features. Also, the performance of age prediction algorithms generally decreased as the number of categories predicted increased. However, trends in the literature suggest that multi-category or numeric age predictions may improve as more advanced algorithms are used. Likewise, in the prediction of race, user postings provided valuable information about the user but is best if considered in conjunction with other user metadata. Also, similar to the use of first names as a gender prediction feature, last name appears to be a strong predictor of user ethnicity. The improvement in race and ethnicity prediction based on additional data from users' networks and posting could be attributed to within-network segregation by race or ethnicity.

### Measures of Performance

Several of the studies included in this review did not evaluate prediction performance. Nine gender prediction approaches reviewed provided no performance metrics, and all of these used profile metadata to generate predictive features. Similarly, only one of the seven age prediction approaches and none of the six race/ethnicity prediction approaches that relied only on profile metadata reported performance metrics. Furthermore, of the studies reporting a performance metric, most used accuracy. However, accuracy is not always a good measure of performance, particularly for studies with a class imbalance because the model may be over-trained to recognize the majority group, and the overall accuracy may be excessively weighted by classification performance specific to this group [134, 135]. To assess how well these prediction methods are performing, accuracy needs to be reported alongside other measures of performance such as, precision, recall, F1-score, or AUC.

**Data Limitations**

While further work is needed to explore the potential value of profile-based prediction features (such as, user description and user names), findings from this review suggest that users' postings contain valuable information for predicting users' age and race/ethnicity. This has important implications for the scalability of this work. While sites such as Twitter and Instagram allow researchers to collect information via public Application Programming Interfaces (APIs), these applications are typically rate limited and may limit the scalability of a particular method [69]. For instance, when gathering timelines, the Twitter API can only make 900 calls per 15-minute window. After reaching a rate limit, the API requires 15 minutes to reset. Although these rate limits may not be a problem when gathering data for several thousand users, gathering data on hundreds of thousands or millions of users would be time consuming. For researchers studying time-sensitive topics – such as infectious disease outbreaks – this delay may be prohibitive. Furthermore, collecting, storing and processing hundreds of tweets or ties per user for very large samples is computationally intensive and may not be accessible to applied researchers with limited resources.

Further, several studies did not use individual ground truth data in the evaluation of their methods. The manner in which the test and training data used are collected may impact the results of proposed prediction methods. The studies evaluated engaged in a variety of techniques for measuring users' 'true' age, race and gender. Some relied on self-reported information contained within the user's postings – for instance, by collecting recent tweets in which a user mentions their current age [76], states that it is their birthday [56, 92, 96], or describes their racial or ethnic identity [54, 72]. Some studies use hand-coded evaluations of users' profiles and/or other online presences [85, 87, 89] to assess demographics, and some use facial recognition software such as Face++ [115] or crowdsourcing platforms such as Amazon's

Mechanical Turk [78] to generate ground truth measures. While some studies (e.g., Bergsma and colleagues [48]) address the effect of differently collected training datasets, further work is needed to assess these potential biases. Ground truth data on social media and blog users, while difficult to obtain, is important for assessing the validity of a proposed approach.

**Ethical Considerations**

We acknowledge that detection of user demographics poses unique ethical questions for researchers. First, users who provide data for the development and application of classification methods are not treated as "human subjects" in a traditional sense. According to the Belmont Report and Common Rule, social researchers are supposed to exercise justice (i.e., ensuring fair distribution of benefits and risks of research, as well as fair selection of research subjects), beneficence (i.e., an obligation to maximize benefits and minimize harm toward research subjects) and informed consent (i.e., disclosure of objectives and possible consequences of the research toward the subject and society at large)[136]. However, implementing procedures such as informed consent is difficult when working with tens or hundreds of thousands of respondents. Additionally, standards regarding data ownership and publishing typically dictate how researchers may use data from digital sources. Further work is need to determine how to preserve human subjects research standards when working with large, digitally generated datasets [137–139].

Ethical concerns regarding the automatic detection of user characteristics extend beyond the individual. The term 'group privacy' refers to demographic de-anonymization, or the placement of users within specific demographic categories [140–142]. While ethical standards regarding the protection of research subjects typically assume that individual protection leads to group protection [142], research suggests that disclosed affiliation with a particular demographic group may lead to unfair targeting by marketing agencies, political campaigns, and more. For instance, Zwitter [140] notes that political campaigns have used bot accounts to target and engage in fabricated grassroots debates with Twitter users who fit the demographic profile of a particular party's constituency – a practice sometimes referred to as "astroturfing". Researchers seeking to predict user demographics not explicitly shared through their profiles should be aware of the risks that disclosure of group status may generate prior to using this inferred data for analysis or intervention.

## CONCLUSIONS

We found that gender is heavily embedded within users' online presence, and that features extracted from users' profiles – such as first names – may be used to accurately predict this trait. Other demographic traits such as, age and race/ethnicity are more challenging to predict. Furthermore, there is a need to assess approaches relying exclusively on profile metadata with the same rigor as approaches that use information from networks or users' postings. Future work in this area should be mindful of the scalability of proposed methods, as well as the implications of these methods for individual and group privacy.

## ACKNOWLDGEMENTS

This work was funded by the Robert Wood Johnson Foundation (grant #73362).